\documentclass[usenatbib,usegraphicx]{mn2e}
\def\mnras{MNRAS}
\def\apj{ApJ}

\def\Tg{T_g}
\def\Tc{T_\gamma}
\def\Ts{T_s}
\def\Tst{T_{\star}}
\def\Tb{T_b}
\def\nH{n_{\rm H}}
\def\omb{\Omega_b h^2}
\def\co{C_{10}}
\def\ao{A_{10}}
\def\dH{\Delta_{\rm H}}
\def\ds{\Delta_{s}}
\def\dg{\Delta_g}
\def\n{{\bf n}}
\def\bT{\bar{T}}
\begin{document}
\title{The CMBR fluctuations from  HI perturbations prior to
  reionization} 
\author[S. Bharadwaj and S. S. Ali]{Somnath Bharadwaj\thanks{Email:
    somnath@cts.iitkgp.ernet.in} and Sk. Saiyad Ali\thanks{Email:
    saiyad@cts.iitkgp.ernet.in}
\\ Department of Physics and Meteorology\\ and
\\ Centre for Theoretical Studies \\ IIT Kharagpur \\ Pin: 721 302 ,
India }
\maketitle
\begin{abstract}
\citet{Lz} have recently proposed that observations of
the CMBR brightness temperature fluctuations produced 
by the HI inhomogeneities prior to reionization hold the promise of
probing the primordial power  spectrum  to a hitherto unprecedented
level of accuracy. This requires a precise quantification of the
relation between density perturbations and brightness temperature
fluctuations. Brightness temperature fluctuations arise from two
sources (1.) fluctuations in the spin temperature, and (2.)
fluctuations in the HI optical depth, both of which are caused by
density perturbations. For the spin temperature, we investigate in
detail its evolution  in the presence of HI fluctuations. For the
optical depth, we find that it is affected by density perturbations
both directly and through  peculiar velocities which move
the absorption features around in frequency. The latter effect, which
has not been included in earlier studies,   is
similar to the redshift space distortion seen in galaxy surveys and
this can cause changes of $50 \%$ or more in the birghtness
temperature fluctuations. 
\end{abstract}
\section{Introduction}
The possibility of probing the universe at high redshifts using the
HI $21 \, {\rm cm}$ line has been the topic of extensive theoretical 
investigation. 
This is  perceived to be the most promising
window for studying the ``dark ages'',  the era between the decoupling
of the CMBR from the primeval plasma at $z \sim 1000$ and the
formation of the first luminous objects at $z\sim 20$
(\citealt{hogan}; \citealt{scott}; \citealt{madau};
 \citealt{tozzi}; \citealt{barkana}; \citealt{isfm}; \citealt{miralda})
After decoupling    
the gas temperature $\Tg$ is maintained at CMBR temperature $\Tc$
through collisions of the CMBR photons with the small fraction of 
electrons that survive the process of recombination. The
collision process becomes ineffective in coupling $\Tg$ to $\Tc$ at $z
\sim 200$. In the absence of external heating  at $z<200$ the gas
cools adiabatically with  $\Tg \propto (1+z)^2$ while  
$\Tc \propto (1+z)$. The spin temperature $\Ts$ is strongly
coupled to $\Tg$ through the collisional spin flipping process until
$z \sim 70$. The collisional process is weak at lower redshifts, and
$\Ts$ again approaches $\Tc$.  This gives a range of  redshifts 
where  $\Ts < \Tc$. We then have a window in redshift $30 \le z \le
200$, or equivalently in  frequency  $\nu =1420 {\rm MHz}/(1+z)$ where
the HI will produce absorption features in the CMBR spectrum. 

In a recent paper \citet{Lz} propose that observations of the angular
fluctuations in the CMBR brightness temperature 
$\Tb$  arising from the HI absorption can be used to study the power
spectrum  of density fluctuations at small scales to a level of accuracy
far exceeding those achievable  by  any other means. 
The enormous wealth of information arises from the fact that
observations at different frequencies which are sufficiently separated 
will provide independent estimates of the power spectrum at the
same wave number $k$. These observations will probe the power spectrum
before the epoch of structure formation,  and they hold the  possibility
of revealing the entire primordial power spectrum down to very small
scales.  In another recent paper \citet{gnedin} have studied the
linear fluctuations in the 21 cm emission from the pre-reionization
era. They show that that it should be possible to detect this
signal against the foreground contaminations in the frequency
domain. This signal is expected to constrain the equation of state of
the universe at high $z$. 

The CMBR brightness temperature is related to $\Ts$ and the HI number 
density $\nH$ as  $\Tb \propto (1-\Tc/\Ts) \,\nH$. Fluctuations in
$\Tb$ arise from fluctuations in $\nH$  directly and also through
fluctuations in $\Ts$ which in turn arise from fluctuations in
$\nH$. In calculating  the fluctuations in $\Ts$, \citet{Lz} consider
only one process,  namely  the change in the collision 
rate arising from fluctuations in $\nH$. Perturbations  in
$\nH$ will also produce perturbation in $\Tg$, which in turn will
affect $\Ts$. This effect has not been taken into account in their
work.

Density perturbations produce peculiar velocities. 
This causes the frequency of the HI absorption features to be 
shifted by the line of sight component of the peculiar velocity.   
This effect will rearrange the HI absorption features in frequency
space where converging velocity patterns  appear as enhancements in
the HI number density and diverging velocity patterns appear as
decrements in the HI number density.  It may be noted that this is the
familiar linear redshift space distortion (Kaiser effect,
\citealt{kais}) seen in galaxy redshift surveys. This effect has been
studied by \citet{bharadwaj1} in the context of cosmological HI
emission  from $z\sim 3.5$. 

It is important to identify and take into account all possible
contribution to the brightness temperature fluctuations, if these are
to be used to extract precise information about the power spectrum and
the equation of state of the universe. In this paper we study two 
effects which will contribute to brightness temperature fluctuations,
namely (1.) perturbations in the gas temperature produced by density
fluctuations, and (2.) the effect of redshift space distortions. 
To the best of our knowledge, these effects have not been included in
earlier works. 

We next present an outline of the paper. In Section
2. we discuss the  processes involved in determining the brightness
temperature fluctuations and present the relevant equations. In
Section 3, we present our results and discuss their consequences. It
may be noted that we have used $(\Omega_{m0}, \Omega_{\Lambda0},
\Omega_{b0} h^2, h) = (0.3,0.7,0.02,0.7)$ whenever  specific values
have been needed for the cosmological  parameters. 

\section{Calculating the brightness temperature fluctuations.}
\begin{figure}
\includegraphics[width=84mm]{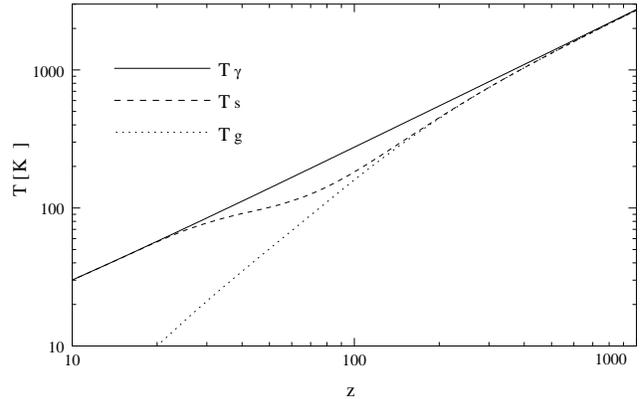}
\caption{This shows the evolution of the CMBR temperature, the gas
  temperature and the spin temperature as the universe expands.}
\label{fig:a1}
\end{figure}

We first consider the evolution of the gas temperature  after the
recombination era $(z \sim 1000)$ when it becomes largely  neutral.
This is governed by the equation 
\begin{equation}
\frac{\partial \Tg}{\partial z}-\frac{2 \Tg}{3 \nH} \frac{\partial
  \nH}{\partial z} 
=\frac{-9.88\times 10^{-8}}{\omb }\, (1+z)^{3/2} (\Tc-\Tg)
\label{eq:a1}
\end{equation}
The third term in the above equation represents the energy transfer
from the CMBR to the gas through the collisions with the residual
electrons \citep{peebles}. This terms tries to maintain the gas
temperature at the 
CMBR temperature as the universe  expands. 
The second term is the change in $\Tg$ in adiabatic expansion. 
If the HI is uniformly distributed, then $\nH \propto (1+z)^3$ 
and in the absence of the CMBR heating we have $\Tg \propto (1+z)^2$. 
Collisions are able to maintain $\Tg =\Tc =2.73 \, {\rm K} \, (1+z)$
up to a 
redshift $z \sim 200$ (Figure \ref{fig:a1}) after which  $\Tg \propto
(1+z)^2$. 

We next consider the evolution of the spin temperature $\Ts$ which is
defined through the relation  
\begin{equation}
\frac{n_1}{n_0}=\frac{g_1}{g_0} e^{-\Tst/Ts}
\end{equation}
where $n_1$, $n_0$ are the population densities and $g_1=3$, $g_0=1$  
the spin  degeneracy factors of the excited and the ground states
of the HI 21 cm transition. In this equation   $\Tst=h_p \nu_e/k_B=
0.068 {\rm   K}$, where $h_P$ is the Planck's constant, $\nu_e=1420
{\rm MHz}$ is the frequency corresponding to  the 21 cm line and $k_B$
is  Boltzmann's constant. 
The evolution of the ground state
population  density is governed by two processes, one collisional and
the other radiative 
\begin{eqnarray}
\left( \frac{\partial}{\partial t} + 3 \frac{\dot{a}}{a} \right) n_0
&=&n_1 \co  -n_0 C_{01} \nonumber \\  
&+&  n_1 \ao + (n_1 B_{10} - n_0 B_{01}) \, I_{\nu_e} 
\label{eq:a3}
\end{eqnarray}
where $a(t)$ is the scale factor, $C_{01}$ and $C_{10}$ are the
collisional excitations and de-excitation rates of the hyperfine
levels, $\ao$ is the Einstein 
spontaneous emission coefficient, $B_{01}$ and $B_{10}$ are the
Einstein $B$ coefficients and $I_{\nu_e}$ is the specific intensity of
the background radiation at $\nu_e$.

In the regime of our interest $\Ts,\,  \Tg, \, \Tc \gg \Tst$ and we
can use the approximation $e^{-\Tst/T} = 1 - (\Tst/T)$ throughout. 
Also, the fact that in equilibrium the collisional processes and the 
radiative process are separately balanced gives us the relations 
$C_{01}=3 (1- \Tst/\Tg) \co$ and  $B_{01}=3 B_{10}=(3 \lambda_e^3/ 2
h_p c) \ao$ where $\ao=2.85 \times 10^{-15} s^{-1}$
\citep{rybicki}. The collisional 
de-excitation rate can be written as  $\co=(4/3) \kappa(1-0) \nH$
where the values of $\kappa(1-0)$ is tabulated as a function of
$\Tg$ \citep{allison}. Using these and equation (\ref{eq:a3}) we
obtain an equation 
for the redshift evolution of $\Ts$
\begin{eqnarray} 
&&\frac{\partial}{\partial z} \left(\frac{1}{\Ts} \right) =
-\frac{4}{H(z) (1+z)} \times  \nonumber \\  && \hspace{1.1cm}
\left[  \left(  \frac{1}{\Tg}- \frac{1}{\Ts} \right)  \co 
+   \left(  \frac{1}{\Tc}- \frac{1}{\Ts} \right) \frac{\Tc}{\Tst} \ao 
\right]
\label{eq:a4}
\end{eqnarray}
where $H(z)$ is the Hubble parameter. The collisional  term tries to set
the spin temperature at the same value as the gas temperature while
the radiative term tries to set it at the CMBR temperature, which
process dominates being decided by the rate coefficients. At high
redshifts the collisional process dominates and the spin temperature
closely follows the gas temperature. At lower redshifts  $\nH$ falls 
substantially, the collisional process looses out to the radiative
process and the spin temperature approaches the CMBR temperature. 
Figure \ref{fig:a1} shows the evolution of the spin temperature as the
universe expands. 

We next shift our attention to the effect of HI density perturbations 
$\dH(\bf{x},z)=(\nH(\bf{x},z)-\bar{\nH}(z))/\bar{\nH}(z)$. These will
produce fluctuations in the gas temperature. If the gas were
undergoing adiabatic expansion, the fluctuations in the gas
temperature $\dg(\bf{x},z)=(\Tg(\bf{x},z)-\bar{\Tg}(z))/\bar{\Tg}(z)$
would be related to $\dH$ through $\dg=(2/3) \dH$. This will get
modified because of the energy that is pumped into the gas from 
the CMBR and $\dg=0$ during the era when $T_g=\Tc$. We define a
function $g(z)=\frac{\partial\dg}{\partial \dH}$.  such that $\dg(z)=g
\, \dH(z)$. Using this in equation (\ref{eq:a1}) we obtain 
\begin{equation}
\frac{d g}{dz}=
\frac{9.88\times 10^{-8} \, \Tc}{\omb \, \Tg }\, (1+z)^{3/2} \,  g +
(\frac{2}{3}-g) \, \frac{1}{\dH} \frac{\partial \dH}{\partial z} 
\label{eq:a5}
\end{equation}
The first term in RHS arises from the coupling of the gas to the CMBR
and it 
tries to set  $g(z) \rightarrow 0$ while the second term corresponds
to adiabatic expansion and it tries to make $g(z) \rightarrow
2/3$. The quantity $g(z)$ is expected to evolve from $g(z)=0$ to
$g(z)=2/3$ as the universe expands and the contribution of the heat
pumped into   the gas goes down. 

\begin{figure}
\includegraphics[width=84mm]{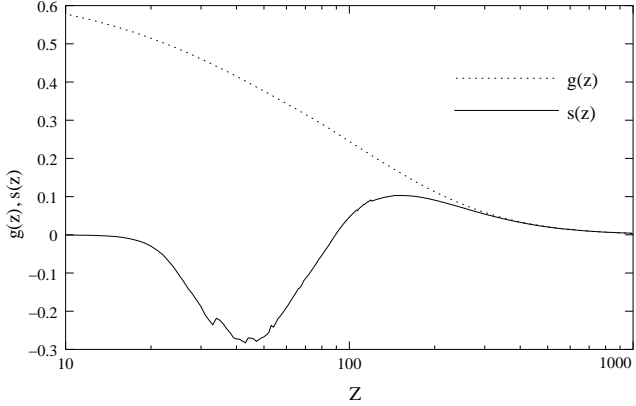}
\caption{This shows the evolution of the functions $g(z)$ and $s(z)$
  defined in the text.}
\label{fig:a2}
\end{figure}

An interesting feature is that $g(z)$ depends on the growth rate of
density fluctuations. For example, it follows from equation
(\ref{eq:a5}) that a static density perturbation which does not evolve
in time will not produce fluctuations in the gas  temperature.  Here
we  assume that $\dH$ follows the dark matter perturbation and
grows as $\dH \propto a(z)$.  The result of integrating equation
(\ref{eq:a5}) is shown in figure \ref{fig:a2}. We see that $g(z) \sim 0.3$ 
in the redshift range of our interest ($30 \le z \le 100$). 

We finally come to the fluctuations in the
spin temperature 
$\ds(\bf{x},z)=(\Ts(\bf{x},z)-\bar{\Ts}(z))/\bar{\Ts}(z)$
produced by density perturbations. From equation
(\ref{eq:a4}) we see that fluctuations in $\Ts$ can arise from
changes in $\Tg$ and from changes  in the collision rate.  The  changes
in  collision rate 
$\co=(3/4) \kappa(1-0) \nH$  will come about directly through changes  
in $\nH$ and also through changes in $\Tg$ which will affect the value
of $\kappa(1-0)$. Taking into account both these effect we have
$\Delta \co = (1+(2/3)\,  d ln \kappa/d ln \Tg)\,  \co \,
\dH$. Defining  a function
$s(z)=\frac{\partial\ds}{\partial \dH}$.  such that $\ds(z)=s \,
\dH(z)$ and  using equation (\ref{eq:a4}) we obtain 
\begin{eqnarray}
&& \frac{d s}{dz}=-s \frac{1}{\dH} \frac{\partial \dH}{\partial z} +
  \frac{4}{H(z)(1+z)} \left[ \left\{ \frac{\Ts}{\Tg}(s-g) \right. \right.
\nonumber  \\
&& \left. \left. +\left( \frac{\Ts}{\Tg}-1
  \right) \left(1   + \frac{d \ln \kappa}{d \ln \Tg}\right) \right\}
  \co + s    \frac{\Ts}{\Tst} \ao \right]  \,
\end{eqnarray}
Here again, the evolution of $s(z)$, like that of $g(z)$, depends on
the time evolution the density fluctuations.  Figure \ref{fig:a2}
shows the evolution of $s(z)$ under the assumption $\dH \propto a(z)$.  
We find that $s(z)>0$ for $z>90$ {\it ie.} a positive density
perturbation causes the spin temperature to go up, and  the effect is
opposite at $z < 90$. 

During the era when $\Ts < \Tc$  the HI along  a  line of sight $\n$
reduces the CMBR  brightness temperature at the frequency $\nu$ by an
amount  
\begin{equation}
\Tb(\n,\nu)=(\Ts - \Tc) \, \tau/(1+z)\,.
\label{eq:a7}
\end{equation}
Here $\tau$ is the optical depth of the 21 cm HI  
absorption given by 
\begin{equation}
\tau=\frac{3 \, \nH \, h_P \, c^2 \,  \ao \, a^2(z)}{32 \, \pi \, k_B
  \Ts    \nu_e} 
\mid \frac{\partial r}{\partial \nu} \mid
\label{eq:a8}
\end{equation}
where $r$ is the comoving distance to the HI whose 21 cm absorption is
redshifted to $\nu$.  

We are interested in the angular fluctuations of the brightness
temperature $ \Tb(\n,\nu)$. HI density fluctuations will produce
fluctuations in $ \Tb(\n,\nu)$  through the fluctuations in the spin
temperature discussed earlier. Density fluctuations will also directly
affect $\Tb$  through variations in the  optical depth which we now
calculate. The relation between the comoving distance $r$ and the
frequency $\nu$ is given by
\begin{equation}
r= \int^{1}_{\frac{\nu}{\nu_e (1-v/c)}} \hspace{.2cm} \frac{c \, da}{a^2
  \, H(a)} 
\label{eq:a9}
\end{equation}  
where $v$ is the line of sight component of the peculiar velocity of
the HI which produces the absorption. Density perturbations will, in 
general,  be accompanied by  velocity perturbations and these will move
around the HI absorption features in frequency. Here we assume that 
the HI traces the dark matter and that we can use linear perturbation
theory to relate the peculiar velocity to the density perturbations. 
Incorporating the effect of both the density fluctuations and the
peculiar velocity we have 
\begin{equation}
\tau=\frac{3 \, \bar{\nH} \, h_P \, c^3 \ao}{32 \, \pi \, k_B \Ts
  \nu^2_e H(z)} 
\left[1 + \dH - \frac{1}{H(z)\, a(z)} \frac{\partial v}{\partial r} 
  \right]   \,.
\label{eq:a10}
\end{equation}
Here we have dropped terms of order $v/c$ in the final
expression. Also, we have retained terms only to linear order in
$v$. There is also the effect of our own motion which we have not
included. These effects are not expected to be important. The term
involving the derivative of the peculiar velocity is the dominant
effect, particularly at the small scales of interest here. 

Combining  the effects of the fluctuations in the  optical depth and
in the spin temperature we can write the fluctuations in the CMBR
brightness temperature as 
\begin{equation}
\delta \Tb(\n,\nu)=\bT \left[ \left(1 - \frac{\Tc}{\Ts}\right) \left( 
 \dH - \frac{1}{H \, a} \frac{\partial v}{\partial r}
  \right)  +  \frac{\Tc}{\Ts} s \dH \right]
\end{equation}
where 
\begin{equation}
\bT=2.67 \times 10^{-3} \, {\rm K} \hspace{0.2cm} \frac{\omb}{0.02}  \, 
\frac{(1+z)^{1/2}}{\Omega_{m0}^{1/2} h} 
\end{equation}

It is  convenient to express this in Fourier space where 
\begin{equation}
\dH({\bf x},z)=\int \frac{d^3 k}{(2 \pi)^3} e^{-i {\bf k} \cdot {\bf
    x} }
\Delta({\bf k},z) 
\end{equation}
and the Fourier transform of the peculiar velocity  is given by
  ${\bf v}({\bf k},z)=-i H(z) a(z)  {\bf k} \Delta(\bf{k},z) /k^2$.
Using this we express the fluctuations in brightness temperature as 
\begin{eqnarray}
\delta \Tb(\n,\nu)&=&\bT \int \frac{d^3 k}{(2 \pi)^3} e^{-i k r \mu}
\Delta({\bf k},z) \times \nonumber \\ & &
 \left[ \left(1 - \frac{\Tc}{\Ts}\right) \left(1+\mu^2 
  \right)  +  \frac{\Tc}{\Ts} s \right]
\end{eqnarray}
where $\mu$ is the cosine of the angle between the comoving wave
vector  ${\bf k}$ and the line of sight $\n$. 

We now calculate the angular power spectrum of the brightness
temperature fluctuations resulting from the density fluctuations
$\Delta({\bf k},z)$. The statistical properties of $\Delta( {\bf k},z)$
are specified  through the 3D  power spectrum  defined as 
\begin{equation}
\langle \Delta({\bf k},z) \Delta({\bf k^{'}},z) \rangle = (2 \pi)^3
\delta_{D}^3({\bf k-k^{'}})\,  P(k,z)
\end{equation}
where $\langle... \rangle$ denotes ensemble average and $\delta_{D}()$
is the Dirac delta function. 

The angular power spectrum is calculated  by decomposing the angular
dependence of $\delta \Tb$ into 
spherical harmonics with expansion coefficients $a_{lm}(\nu)$ and
using these to calculate 
the angular power spectrum $C_l(\nu)=\langle \mid a_{lm}\mid^2
\rangle$. The angular power spectrum can be expressed in terms of the
3D power spectrum as 
\begin{eqnarray}
C_l(\nu)=4 \pi \bT^2 \int \frac{d^3 k}{(2 \pi)^3} && P(k,z)
\hspace{.3cm} 
\left[ \left(1 - \frac{\Tc}{\Ts}\right) J_l(k r) \right. \nonumber \\
&& \left.
 +  \frac{\Tc}{\Ts} \, s \,   j_l(k r)   \right]^2 
\label{eq:a16}
\end{eqnarray}
where $j_l(k r)$ are the spherical Bessel functions and 
\begin{eqnarray}
J_l(k r)&=&\left[-\frac{l (l-1)}{4 l^2-1} j_{l-2}(k r)
+ \frac{2(3 l^2 + 3 l -2)}{4 l^2 + 4l -3} j_l(k r)
  \right. \nonumber \\
&&     \left. 
 -\frac{(l+2)(l+1)}{(2l+1)(2l+3)}
  j_{l+2}(kr) \right] 
\end{eqnarray}
 \section{Results and Discussion.}
In this paper we have investigated in detail the CMBR fluctuations
produced by HI priot to the epoch of reionization. As proposed by Loeb
and Zaldarriaga (2003), this holds the promise of allowing the
power-spectrum of density fluctuations to be probed to a high level of
precision.  

\begin{figure}
\includegraphics[width=84mm]{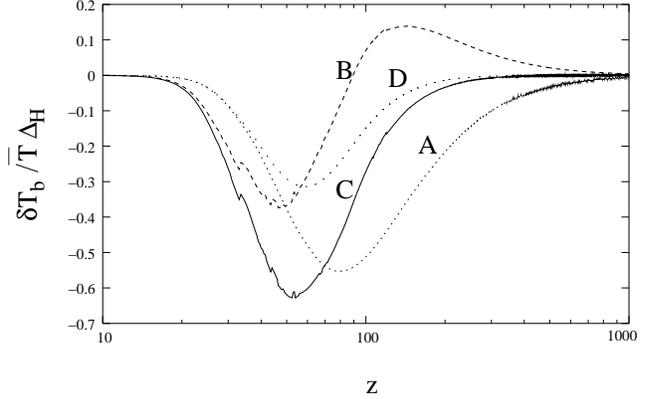}
\caption{This shows how the CMBR brightness temperature fluctuations
  (in units of $\bar{T}$) respond to HI density perturbations. A. shows
  $(1-\Tc/Ts)$ which is how  it responds through changes in the
  optical depth, ignoring the effect of peculiar velocities. Peculiar
  velocities will enhance this effect. B. shows $s \, \Tc/\Ts$ which
  is   how it responds through changes in the spin
  temperature. C. shows the total response for $\mu^2=2/3$. D is the
  same as B without the gas temperature fluctuations.} 
\label{fig:a3}
\end{figure}

HI density perturbations produce fluctuations in the decrement of the
CMBR brightness 
temperature by two means (1.) through fluctuations in the optical
depth, and (2.) through fluctuations in the spin temperature. 
The effect of changes in the optical depth in response to a
positive density perturbation (curve A of  Fig. \ref{fig:a3}) is such
that it  reduces $\Tb$  and enhances the decrement in the
brightness temperature. This effect is maximum at $z \sim 80$. This 
effect is  enhanced by  peculiar velocities. 

The  change in brightness temperature produced by  HI density
perturbations through changes in the spin temperature varies with
$z$ (curve B of Fig. \ref{fig:a3}). Density perturbations increase the
spin temperature and the 
brightness temperature in the redshift range $z > 100$.  
Here the collisional process is very efficient  and
$\Ts$ closely follows $\Tg$. A positive density perturbation 
increases  $\Tg$  which  causes $\Ts$ to also increase.  The effect of
density perturbations on the brightness  
temperature acting through changes in the optical depth and through 
the spin temperature have  opposite signs. The effect of changes 
in the optical depth is larger and the brightness temperature
decrement is enhanced by a positive density perturbation (curve C of
Fig. \ref{fig:a3}).   

In the redshift range $z<100$  positive density 
perturbations lower the spin temperature which enhance the brightness
temperature decrement. In this regime the collisional  process slowly 
looses out to the radiative process and $\Ts \rightarrow \Tc$. A
positive density perturbation enhances the collision rate which pulls
the spin temperature down toward the gas temperature. The two
processes  which contribute toward brightness 
temperature fluctuations both act to enhance the decrement. Curve C of
Fig. \ref{fig:a3} shows the combined effect of both these processes
for the value $\mu^2=2/3$.   We find that the
response of the brightness temperature to density perturbations peaks
at $z \sim 55$. This is somewhat smaller than the value obtained by
\citet{Lz}.  Curve D of Figure 3  shows the contribution to
brightness temperature fluctuations arising from the spin
temperature if the effect of density perturbations on the gas
temperature is not taken into account (eg. \citealt{Lz}). We find that
including the gas temperature  makes a significant change, 
particularly at $z>100$ where there is a qualitative difference
between curves B and D. 

\begin{figure}
\includegraphics[width=84mm]{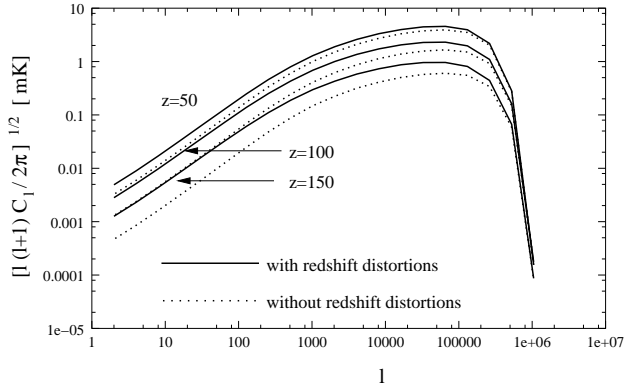}
\caption{This shows the angular power spectrum of the CMBR brightness
  fluctuations at various redshifts with and without the effects of
  peculiar velocities.}
\label{fig:a4}
\end{figure}

We have calculated the angular power spectrum $C_l(\nu)$ of the
brightness 
temperature fluctuations for the COBE normalized ${\rm \Lambda CDM}$
model \citep{bunn}.  The power spectrum has been
suppressed beyond the arbitrarily chosen value $k=14 \, h\, {\rm
  Mpc}^{-1}$  using a Gaussian cut-off.  Our
results are in qualitative agreement with  \citet{Lz}. We find that
the signal peaks at $z \sim 50$ (Fig. \ref{fig:a4}) where the
product of the growing 
mode of density perturbations and the response of brightness temperature
to density perturbations is maximum. 

 To gauge the effect of  peculiar
velocities, we have calculated  $C_l(\nu)$ without incorporating  this
effect. This is easily done  
by replacing $J_l(kr)$ with $j_l(k r)$ in equation (\ref{eq:a16}). The
results are  shown in Fig. \ref{fig:a4}. We find that the peculiar
velocities increase $\sqrt{C_l}$  by more than $50 \% $.  

We have also quantified the effect of gas temperature fluctuations. 
We find that for $z < 100$   the values of $\sqrt{C_l}$ are
$\sim 10 \%$ lower if the gas temperature is not taken into account,
and   the effect is reversed at $z>100$ where  $\sqrt{C_l}$ is $\sim 30 \%$
higher.

The ability to probe the dark matter power spectrum using the
$C_l(\nu)$s will be restricted to scales $k < k_{\rm J}$ where $k_{\rm
  J} $ is the Jeans wave number. This has a nearly constant value
$\sim 500 \,  h \, {\rm Mpc}^{-1}$ in the redshift range of interest.  
The HI power spectrum on 
scales smaller than the Jeans length scale  is  interesting in its
own right.  The HI perturbations will undergo acoustic
oscillations  on these scales. The spin temperature
fluctuations and the peculiar velocities produced by density
perturbations will be quite different from the situation considered
here. On scales slightly larger than the Jeans lengthscale ($2
\pi/k_J$) the density fluctuations are mildly non-linear with
rms. values in the range $.7 > \sigma > .1$,  and  the Z'eldovich
approximation may give  a better description (eg. \citealt{hui}).  

The low frequency cut-off imposed by the earth's
ionosphere restricts observations to frequencies more than $\sim 10 -
20 \, {\rm MHz}$.  
Extracting the HI signal from the contaminations arising from the
galactic and extragalactic foregrounds is going to be a big
challenge. The foregrounds are mostly continuum sources whose
contribution varies slowly with frequency. It will be necessary to
combine both the angular fluctuations and the frequency domain
properties of the CMBR brightness temperature fluctuations in order to
detect it (eg. \citealt{shaver}, \citealt{dmat1}, \citealt{oh},
\citealt{dmat}). 

\section*{Acknowledgments}
SSA would like to acknowledge financial support through a junior
research fellowship of the Council of Scientific and Industrial
Research (CSIR), India.

\end{document}